\documentclass[12pt, peerreview, onecolumn]{IEEEtran}

\usepackage{epstopdf}

\usepackage{amsmath,epsfig,verbatim,amsopn,subfigure,color}
\usepackage{cite,xspace}
\usepackage{array,algorithm,algorithmic}
\usepackage{array}
\usepackage{multirow,array}%
\usepackage{multicol}%
\usepackage{setspace}
\usepackage{here}
\usepackage{amssymb}
\usepackage{dsfont}
\usepackage{subfigure}
\usepackage{soul}
\usepackage{algorithm}
\usepackage{breqn}

\usepackage[
    top    = 0.480in,
    bottom = 0.480in,
    left   = 1.4cm,
    right  = 1.4cm]{geometry}

\usepackage[center]{caption}

\usepackage{tikz}
\usetikzlibrary{matrix,fit}
\usetikzlibrary{shapes,arrows}

\usepackage{ graphicx,  bm,  bbding,pifont}
\usepackage{epsf, epsfig}
\usepackage{graphics}
\usepackage{times,relsize}
\usepackage{fancybox,calc}
\usepackage{graphics, color, psfrag,amsfonts}
\usetikzlibrary{decorations.markings}
\usepackage{sidecap}
\usetikzlibrary{shapes,arrows}
\usetikzlibrary{decorations.pathreplacing}
\tikzstyle{decision} = [diamond, draw, fill=white!20,
    text width= 2em, text centered, node distance=1cm, inner sep=2.1pt]
\tikzstyle{block} = [rectangle, draw, fill=white!20,
    text width=2em, text centered, rounded corners, minimum height=2.1em]
\tikzstyle{new_block} = [rectangle, draw, fill=gray!30,
    text width=2em, text centered, rounded corners, minimum height=2.1em]
\tikzstyle{block2} = [rectangle, draw, fill=white!20,
    text width=2em, text centered, rounded corners, minimum height=2.7em]
\tikzstyle{line} = [draw, -latex']
\tikzstyle{state}=[circle, draw, fill=white!10,
    text width=2em, text badly centered, inner sep=2.5pt,minimum height=1.5em]
\tikzstyle{cloud} = [ellipse,fill=gray!35, node distance=3cm,
    minimum height=2em]

\ifCLASSOPTIONonecolumn

\else
    
\fi

\makeatletter
\renewcommand*{\@opargbegintheorem}[3]{\trivlist
 \item[\hskip \labelsep{\itshape #1\ #2}] {\itshape (#3):} {\normalfont}}
\makeatother

\begin{document}
\title{In Order Packet Delivery  in Instantly Decodable Network Coded Systems over Wireless Broadcast}
\vspace{-7mm}
\author{Mohammad S. Karim,  Parastoo Sadeghi, Neda Aboutorab  and Sameh Sorour
\thanks{M. S. Karim, P. Sadeghi and N. Aboutorab  are with the  Australian National University, Australia (\{mohammad.karim,
parastoo.sadeghi, neda.aboutorab\}@anu.edu.au). S. Sorour is with the King Fahd University of Petroleum and Minerals (samehsorour@kfupm.edu.sa).}



}

\maketitle

\maketitle
\vspace{-7mm}
\begin{abstract}
In this paper, we study  in-order packet delivery in instantly decodable network coded systems for wireless broadcast networks. We are  interested in applications, in which  the successful delivery of a packet  depends on the correct reception of this packet and    all its preceding packets. We  formulate the  problem of minimizing the number of undelivered packets to all receivers over all transmissions until completion  as a stochastic shortest path (SSP) problem. Although finding the optimal packet selection policy using SSP is computationally complex, it allows  us to systematically exploit the problem structure and draw guidelines for efficient packet selection policies that can  reduce the number of undelivered packets to all receivers over all transmissions until completion. According to these guidelines, we design a simple heuristic packet selection algorithm. Simulation results illustrate that our proposed  algorithm  provides quicker packet  delivery  to the receivers   compared to the existing algorithms in the literature.
\end{abstract}


\vspace{-5mm}
\vspace{-7mm}
\section{Introduction}
Network coding (NC) has shown great potential to improve  throughput,  delay and a balance between
throughput and delay in  wireless  networks \cite{katti2006xors,wanginstantly,sorour2012completion,aboutorabenabling,sadeghi2010optimal,sorour2010decoding,le2013instantly,li2011capacity}. These merits of NC make it an attractive  candidate  for  numerous applications. In  this paper, we are interested in  applications with in-order packet delivery constraint, where  a packet can be delivered to the application if this packet and  all its preceding   packets are successfully decoded \cite{sundararajan2009feedback}. Examples of such  scenarios are  cloud based applications, Dropbox and Google Drive, where  packets represent instructions that need to be executed in-order.   Furthermore, audio and video streaming  applications,   NetFlix and YouTube, need to    play packets in-order and on-time  in order to prevent interruption  of the stream. In transmission control protocol (TCP),   packets are delivered to the application  in-order and thus, out-of-order packet receptions at the receiver can flood its buffer with undelivered packets.   For such scenarios, it is desirable to design NC schemes so that the received packets are quickly decoded and delivered.

While most of the  NC schemes offer high throughput, they do not necessarily  provide quick decoding and delivery of the received packets. For instance, random linear network coding (RLNC) \cite{Ho2006random} achieves the best throughput  for broadcasting a block of packets, at the expense that  no packet can be decoded and delivered until the receivers collect sufficient number of independent coded packets.  Such delay performance of  RLNC  makes it less attractive to the delay-sensitive applications such as audio and video streaming. In order to reduce the delay of network coded systems, an attractive strategy is to use  \emph{instantly decodable network coding (IDNC)}\cite{katti2006xors,wanginstantly,sorour2012completion,aboutorabenabling,sadeghi2010optimal,sorour2010decoding,le2013instantly,li2011capacity}. IDNC aims to provide instant packet decodability upon successful packet reception at the receivers  and thus,  allows the instant use of the received packets. Moreover, the encoding and decoding processes of IDNC are performed using simple XOR operations.  These simple decoding operations  reduce packet overhead and
are suitable for implementation on mobile devices. In IDNC systems, the immediately undecodable  packets  are  discarded and thus, there is no additional buffer requirements at the receivers to store undecoded  packets.

Due to these desirable properties, the authors in \cite{sadeghi2010optimal,sorour2010decoding,le2013instantly}   considered IDNC to service the maximum number of receivers with a new packet in each transmission. In \cite{sorour2012completion,aboutorabenabling}, the authors addressed the problem of minimizing
the number of transmissions required for broadcasting a block of packets in IDNC systems and formulated the
problem into a stochastic shortest path (SSP) framework. The works in \cite{sorour2012completion,aboutorabenabling,sadeghi2010optimal,sorour2010decoding,le2013instantly} considered the applications, in which each decoded packet brings new information and  is immediately delivered to the application irrespective of its order.   Moreover, the authors in \cite{li2011capacity}  considered video streaming with sequential packet delivery deadlines and showed that, for sufficiently large video files, their IDNC schemes are asymptotically throughput-optimal for the two-receiver and three-receiver systems subject to deadline constraints.

In this paper, inspired by  applications that are delay-sensitive and require in-order packet delivery, we are interested in designing a comprehensive IDNC framework  that can provide contiguous and in-order packet delivery to the
receivers in wireless broadcast networks. In such scenarios,   IDNC schemes need to systematically address the  complicated   interplay of   servicing a set of receivers with the first in-order missing packets and servicing another set of receivers with  other missing packets in each transmission.  In fact, servicing a receiver with any other missing packet can deliver a burst of in-order decoded packets to the application when the first in-order missing packet is decoded in future transmissions.  These aspects of in-order packet delivery constraint lead us to a totally different problem with its own features, problem formulation and solution compared to those in \cite{sorour2012completion,aboutorabenabling,sadeghi2010optimal,sorour2010decoding,le2013instantly}, which ignored in-order packet delivery constraint in IDNC systems.

In the context of this paper, the most related work is \cite{wanginstantly}. In particular,  the authors in \cite{wanginstantly}  discussed the  delivery dependency between source packets with motivating examples  and  designed a heuristic packet selection algorithm that aimed to reduce the number of  transmissions  while respecting in-order packet delivery to the receivers. In contrast, we  represent all feasible packet combinations in IDNC in the form of an IDNC graph and formulate the problem of minimizing the  number of  undelivered packets to all receivers over all transmissions until completion into an SSP framework.  Our SSP formulation is a sequential decision making process in which  the decision is made at each time slot and  takes into account the future situations,  such that the  receivers  are not necessarily always serviced with their  first in-order missing packets but  also serviced with  other missing packets. Although solving this SSP formulation is computationally  complex,  combined with the IDNC graph representation, it allows us to systematically   draw more comprehensive
 guidelines  for  efficient packet selection policies compared to \cite{wanginstantly}. Based on these guidelines, we design  a simple heuristic packet selection  algorithm. Simulation results show that our designed IDNC algorithm  outperforms the IDNC algorithm in \cite{wanginstantly} in terms of quick packet   delivery  to the receivers and number of required transmissions.

\vspace{-5mm}
\section{System Model} \label{tools}
We consider a wireless sender that wants to  deliver a set   of $N$ source packets  $\mathcal{N} = \{ P_1,...,P_N\}$ to a set  of $M$ receivers $\mathcal{M} = \{R_1,...,R_M\}$.\footnotemark \footnotetext{Note that when the
context is clear, we may  denote packet $P_j$ and receiver $R_i$ by their index values $j$ and $i$, respectively.} All source packets of $\mathcal{N}$ can be delivered to the application of each receiver only in order,  meaning that the successful delivery of a packet to the application  depends on the correct reception of this packet and all its preceding packets. For instance, packet $P_j$ can be delivered to the application only if packets  $P_1,...,P_j$ are  decoded.   Time is slotted and the sender can transmit one packet per  time slot $t$. Each transmitted packet is subject to independent Bernoulli erasure at receiver $R_i, R_i\in \mathcal{M}$, with the probability $\epsilon_i$, which is assumed to be fixed during the  transmission period.  Each receiver listens to all transmissions  and  feeds back to the sender a positive or negative acknowledgement for each received or lost packet.


After each transmission,  the sender stores the reception  status of all packets of all receivers in an $ M\times N $  \emph{state feedback matrix (SFM)} $\mathbf{F} = [f_{i,j}],$ $\; \forall R_i  \in \mathcal{M}, P_j\in \mathcal{N}$ such that:
 \begin{equation}
  f_{i,j} =
   \begin{cases}
    0 & \text{if packet $P_j$ is received by receiver $R_i$}, \\
    1 & \text{if packet $P_j$ is missing at receiver $R_i$}.
   \end{cases}
 \end{equation}

\newtheorem{examples}{\textbf{Example}}
\begin{examples}
An example of SFM with $M = 2$ receivers and $N = 6$ packets is given as follows:
\begin{equation} \label{largeSFM1}
\mathbf{F} = \begin{pmatrix}
  1 & 0 & 1 & 0 & 0 & 0 \\
  0 & 0 & 1 & 1 & 0 & 1\\
\end{pmatrix}.
\end{equation}
\end{examples}

In this paper, a missing packet of a receiver can be one of the following two cases:
\begin{itemize}
\item \emph{Next needed packet}: The missing packet $P_j$  of receiver  $R_i$ is referred to as the next needed packet, if all its preceding packets  (i.e., $P_1,...,P_{j-1}$) have been decoded and delivered to this receiver. In Example 1, packet $P_1$ and  packet $P_3$ are  the next needed packets of receiver $R_1$ and receiver $R_2$, respectively.
\item \emph{Needed packet}: A missing packet of receiver $R_i$, except the next needed packet, is referred to  as a needed packet of this receiver. In Example 1, packets $P_4$ and $P_6$ are needed packets of receiver  $R_2$.
\end{itemize}

Based on the SFM, four sets of packets can be attributed to each receiver $R_i$ at any given time slot $t$:
\begin{itemize}
\item The \emph{Has} set ($\mathcal{H}_i$) is defined as the set of packets successfully decoded by receiver $R_i$.
\item The \emph{Wants} set ($\mathcal{W}_i$) is defined as the set of  missing packets at receiver $R_i$. In other words, $ \mathcal{W}_i = \mathcal{N}\setminus \mathcal{H}_i$. In Example 1, the Wants sets of receivers $R_1$ and $R_2$ are $\mathcal{W}_1 = \{P_1, P_3\}$ and $\mathcal{W}_2 = \{P_3, P_4, P_6\}$, respectively.
\item The \emph{Undelivered} set ($\mathcal{U}_i$) is defined as  the set of undelivered packets to receiver $R_i$, which includes  the next needed packet  and all its succeeding packets. In Example 1, the Undelivered sets of receivers $R_1$ and $R_2$ are $\mathcal{U}_1 = \{P_1,P_2,P_3, P_4,P_5,P_6\}$ and $\mathcal{U}_2 = \{P_3, P_4,P_5,P_6\}$, respectively.
\item The \emph{Potential} set ($\mathcal{L}_i$)  is defined as the set of  packets that will be immediately delivered to  receiver $R_i$ upon  decoding  the next needed packet. This set includes all the  packets from the next needed packet   to the following missing  packet.  In Example 1, the Potential sets of receivers $R_1$ and $R_2$ are $\mathcal{L}_1 = \{P_1, P_2\}$ and $\mathcal{L}_2 = \{P_3\}$, respectively.
\end{itemize}
The cardinalities of $\mathcal{H}_i, \mathcal{W}_i, \mathcal{U}_i$ and $\mathcal{L}_i$ are denoted  by $H_i, W_i, U_i$ and $L_i$, respectively  (e.g., $|\mathcal{H}_i| = H_i$).  The set of receivers having \emph{non-empty Wants sets} is denoted by $\mathcal M_w$ (i.e., $\mathcal{W}_i \neq \varnothing, \forall R_i \in \mathcal M_w$). In Example 1,  $\mathcal{M}_w = \{R_1, R_2\}$.  A summary of the main notations used throughout the paper is presented  in Table I.

\newtheorem{definitions}{\textbf{Definition}}
\begin{definitions}
\emph{A transmitted packet is instantly decodable for receiver $R_i$ if it contains one source packet from $\mathcal{W}_i$.}
\end{definitions}
\begin{definitions}
\emph{The completion time is defined as the number of transmissions required to deliver all the packets in $\mathcal{N}$ to all the receivers in $\mathcal{M}$.}
\end{definitions}
\begin{definitions}
\emph{Receiver $R_i$ is targeted by  packet $P_j $ in a  transmission  when this receiver  will immediately decode  missing packet $P_j$ upon  successfully receiving   the transmitted packet.}
\end{definitions}


In this paper, having considered the in-order packet delivery constraint, we adopt a single-phase transmission setting, in which the sender exploits the diversity of received and lost packets at different receivers to transmit uncoded  or coded (XORed) packets from the beginning of the transmission. The transmitted packet will be instantly decoded at a subset of, or all,  receivers. Receivers that  cannot immediately decode a new packet from the received packet discard it. This  transmission process is  continued until all receivers obtain all packets. However, a  two-phase IDNC transmission  setting was  widely considered in the literature   \cite{sorour2012completion,aboutorabenabling,sadeghi2010optimal,sorour2010decoding,le2013instantly}, even the in-order packet delivery based IDNC scheme studied in \cite{wanginstantly}, which has some limitations as we now discuss.  
\vspace{-6mm}
\subsection{Limitations   of the  Two-Phase  Transmission Setting  on In-order Packet Delivery} \label{sec:initial}
In the  \emph{initial (first) phase} of the two-phase transmission setting, the sender  transmits $N$ source packets following the order of the packet indices in an uncoded manner. However, once a packet is lost at a receiver  due to channel  erasure in an initial transmission, the Undelivered set of the receiver will remain unchanged in the remaining initial transmissions. In such a case, the receiver may receive and decode new source packets in the remaining initial transmissions, which cannot be immediately delivered to the application. In general,   the initial  phase (i.e., two-phase transmission setting) limits the packet coding options at the sender and, may result in a large number of undelivered packets to all receivers after each initial transmission. We will further illustrate the limitations of the  two-phase transmission setting  in Section \ref{results}.



\vspace{-5mm}
\section{IDNC Packet Generation} \label{graph}
We describe the representation of all  feasible packet combinations that are instantly decodable by a subset of, or  all,  receivers   in the form of a graph. As illustrated in \cite{sorour2010decoding,sorour2012completion}, the IDNC graph $\mathcal{G(\mathcal V, \mathcal E)}$ is constructed  by first inducing a vertex $v_{ij} \in \mathcal V$ for each packet $P_j\in \mathcal{W}_i, \; \forall R_i\in \mathcal{M}$. Two vertices $v_{ij}$ and $v_{kl}$ in $\mathcal{G}$  are connected (adjacent)  by an edge $e_{ij,kl}\in \mathcal E$, when one of the following two conditions holds. (\textbf{C1}): $P_j = P_l$, the two vertices are induced by the same missing  packet $P_j$ of two different receivers $R_i$ and $R_k$. (\textbf{C2}): $P_j\in \mathcal{H}_k$ and $P_l\in \mathcal{H}_i$, the requested packet of each vertex is in the Has set of the receiver of the other vertex.

Given this graph representation, the set of all feasible packet combinations in IDNC  can be defined by the set of all maximal cliques in $\mathcal{G}$ \cite{sorour2010decoding,sorour2012completion}. The sender can generate a coded  packet for a given transmission by XORing all the source packets identified by the vertices of a  maximal clique (represented by $\kappa$) in $\mathcal G$.  Each receiver can have at most one vertex (i.e., one missing packet) in a maximal clique and  the selection of a maximal clique  $\kappa$  is equivalent to  the selection of a \emph{set of targeted receivers }(represented by $\mathcal{T}(\kappa)$). 

\newtheorem{remark}{\textbf{Remark}}
\begin{remark}
It is possible that a  selected maximal clique $\kappa$ in a transmission includes a set of vertices, which are induced by a set of next needed packets and other needed packets. In this paper, the set of receivers  whose \emph{next needed packets} are included in  $\kappa$ is represented   by  $\mathcal{T}_{\rho}(\kappa)$ and the set of receivers whose  other \emph{needed packets} are included in $\kappa$  is represented  by $\mathcal{T}_{\sigma}(\kappa)$. In fact, $ \mathcal{T}_{\rho}(\kappa) \bigcup \mathcal{T}_{\sigma}(\kappa) = \mathcal{T}(\kappa)$.
\end{remark}



\vspace{-5mm}
\section{Problem Formulation using Stochastic Shortest Path (SSP)} \label{problem formulation}

\begin{figure}[t]
\centering
\tikzset{
    table nodes/.style={
        rectangle,
        draw=black,
        align=center,
        minimum height=7mm,
        text depth=0.5ex,
        text height=2ex,
        inner xsep=0pt,
        outer sep=0pt
    },
    table/.style={
        matrix of nodes,
        row sep=-\pgflinewidth,
        column sep=-\pgflinewidth,
        nodes={
            table nodes
        },
        execute at empty cell={\node[draw=none]{};}
    }
}

\begin{tikzpicture}

\matrix (first) [table,text width=7mm,name=table]
{
1 & 0 & 1 & 0 \\
0 & 0 & 1 & 1\\
};
\begin{scope}[xshift=-6cm, yshift=-3cm]
\matrix (second) [table,text width=7mm,name=table] {
0 & 0 & 1 & 0\\
0 & 0 & 1 & 0\\
};
\end{scope}

\begin{scope}[xshift=-2 cm, yshift=-3cm]
\matrix (second) [table,text width=7mm,name=table] {
0 & 0 & 1 & 0 \\
0 & 0 & 1 & 1 \\
};
\end{scope}

\begin{scope}[xshift=2cm, yshift=-3cm]
\matrix (second) [table,text width=7mm,name=table] {
1 & 0 & 1 & 0 \\
0 & 0 & 1 & 0 \\
};
\end{scope}

\begin{scope}[xshift=6cm, yshift=-3cm]
\matrix (second) [table,text width=7mm,name=table] {
1 & 0 & 1 & 0 \\
0 & 0 & 1 & 1\\
};
\end{scope}

\node        (T){};
\node   [below = .55cm]   (T_a){};
\node   [right = -6cm, below = 2cm]   (T_b){};
\node   [right = -2cm, below = 2cm]   (T_c){};
\node   [right = 2cm, below = 2cm]   (T_d){};
\node   [right = 6cm, below = 2cm]   (T_e){};
\path [line] (T_a) -- node[name=t1] {}  (T_b);
\path [line] (T_a) -- node[name=t2] {}(T_c);
\path [line] (T_a) -- node[name=t3] {} (T_d);
\path [line] (T_a) -- node[name=t4] {} (T_e);

\node (C_a)  [right = 3.5cm, above = 0.1cm]  at (T) {$a_1 = P_1 \bigoplus P_4$};
\node (C_b)  [below = .4cm ]  at (C_a){$a_2 = P_3$};

\node  (A_a) [right = -6cm, below = 3cm] at  (T_a){$\mathbf{u}(s) = [2,2]$};
\node  (A_b) [right = 2.8cm] at  (A_a){$\mathbf{u}(s) = [2,2]$};
\node  (A_c) [right = 2.8cm] at  (A_b){$\mathbf{u}(s) = [4,2]$};
\node  (A_d) [right = 2.7cm] at  (A_c){$\mathbf{u}(s) = [4,2]$};
\node  (A_e) [left = 3cm, above = 0.1cm] at  (T){$\mathbf{u}(s) = [4,2]$};

\node  (O_a) [right = -6.5cm, below = 0.65cm] at  (T_a){$(1-\epsilon_1)(1-\epsilon_2)$};
\node  (O_b) [right = 3cm] at  (O_a){$(1-\epsilon_1) \epsilon_2$};
\node  (O_c) [right = 1.7cm] at  (O_b){$\epsilon_1(1-\epsilon_2)$};
\node  (O_d) [right = 2.5cm] at  (O_c){$\epsilon_1 \epsilon_2$};

\end{tikzpicture}
\caption{State representation, action space   and its possible transitions for action $a_1$} \label{stateTran}
\end{figure}
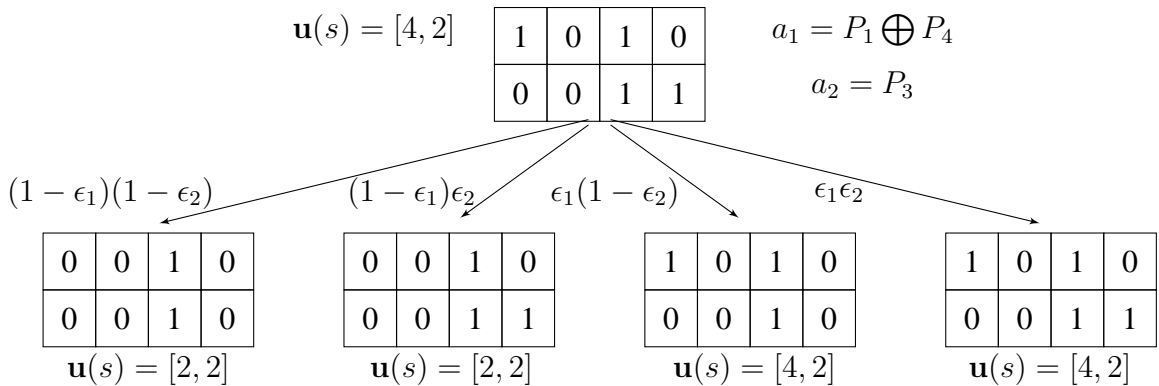

The  problem of minimizing the  number of undelivered packets to all receivers over all transmissions until completion  can be formulated as  a stochastic shortest path (SSP) problem as follows:

\begin{enumerate}
\item \emph{State Space} $\mathcal{S}$: State space $\mathcal{S}$ is  defined by all possibilities of SFM $\mathbf{F}$ and  the Undelivered  sets of the   receivers  resulting from each possible SFM. An SFM of a  state  $s \in \mathcal{S}$ can be represented by  $\mathbf{F}(s)$. Based on  $\mathbf{F}(s)$, we can attribute to  each state $s$   two vectors,  Wants  vector $ \mathbf{w}(s) = [W_1(s),...,W_M(s)]$   and Undelivered vector $\mathbf{u}(s) = [U_1(s),...,U_M(s)]$. Furthermore, we define the  absorbing (i.e., completion)  state $s_a$ as the state in which there is no undelivered packet to any receiver (i.e., $ U_i(s_a) = 0, \forall R_i \in \mathcal{M} $). The size of the  state space is the number of possible variations of SFM, which is $|\mathcal{S}| = O(2^{MN})$.

\item \emph{Action Space} $\mathcal{A}(s)$: The action space $\mathcal{A}(s)$ of  state $s$ consists of the set of  all  possible maximal cliques in the  IDNC graph $\mathcal{G}(s)$, constructed from the SFM $\mathbf{F}(s)$.

\item \emph{State-Action Transition Probabilities}: The state action transition probability $\mathcal{P}_a(s,s')$ for an action $a = \kappa(s) \in \mathcal{A}(s)$ can be defined based on the possibilities of the variations in $\mathbf{w}(s)$ and  $\mathbf{u}(s)$ from state $s$ to its successor state $s'$. To define  $\mathcal{P}_a(s,s')$, we  introduce the following four sets:
    \begin{equation}
    \mathcal{X} = \{ R_i \in \mathcal{T}_{\rho} (\kappa(s))| W_i(s') = W_i(s) -1, U_i(s') = U_i(s) - L_i(s) \}
    \end{equation}
    \begin{equation}
    \mathcal{X}' = \{ R_i \in \mathcal{T}_{\rho} (\kappa(s))|W_i(s') = W_i(s), U_i(s') = U_i(s) \}
    \end{equation}
    \begin{equation}
    \mathcal{Y} = \{ R_i \in \mathcal{T}_{\sigma}(\kappa(s))| W_i(s') = W_i(s) -1 , U_i(s') = U_i(s) \}
    \end{equation}
    \begin{equation}
    \mathcal{Y}' = \{ R_i \in \mathcal{T}_{\sigma}(\kappa(s))| W_i(s') = W_i(s), U_i(s')= U_i(s) \}
    \end{equation}

    Here, the first set $\mathcal{X}$ includes the receivers who have been targeted by their next needed packets and have successfully received the packet. Therefore,  the size of their Wants sets is reduced  by one unit and the size of their Undelivered sets is reduced  by  the size of their Potential sets. The second set $\mathcal{X}'$ includes the receivers who have been targeted by their next needed packets and have lost the packet due to channel erasures. Therefore,  their Wants  and  Undelivered sets  remained unchanged. The third set $\mathcal{Y}$ includes the receivers who have been targeted  by one of their needed packets  and have successfully received the packet. Therefore, the size of their Wants sets is reduced by one unit and their Undelivered sets  remained unchanged. The fourth set $\mathcal{Y}'$ includes the receivers who have been targeted  by  one of their needed packets  and have lost the packet  due to channel erasures.  Therefore, their Wants  and Undelivered sets   remained unchanged.

     Based on the definitions of these four sets, $ \mathcal{P}_{a}(s,s')$  can be expressed as follows:
    \begin{equation}
      \mathcal{P}_{a}(s,s')  = \prod_{i \in \{ \mathcal{X} \cup \mathcal{Y} \} } (1 - \epsilon_{i}). \prod_{i \in \{ \mathcal{X}' \cup \mathcal{Y}' \} }  \epsilon_{i}
    \end{equation}

\begin{examples}
Let us consider  the state representation and the action space  in Fig. \ref{stateTran}. This figure depicts the state-action  transition probabilities and their resulting states when action $a_1$ is selected.
\end{examples}

\item \emph{State-Action Costs}: In the context of  contiguous and  in-order packet delivery,   at  state $s$,   the  expected cost of  action $a$  on each  receiver $R_i \in \mathcal M_w(s)$ can be defined as the expected  number of undelivered packets to  receiver $R_i$ at the successor state $s'$. Now, we  express the expected cost of action $a = \kappa(s) \in \mathcal{A}(s)$ on each receiver $R_i \in \mathcal M_w(s)$  as follows:

   \begin{itemize}
   \item  Consider  receiver $R_i$   has  been targeted by its next needed packet, i.e., $R_i \in \mathcal{T}_{\rho}(a)$. If receiver $R_i$   receives the packet, the size of its  Undelivered set  will be reduced by the size of its  Potential set  (i.e., $U_i(s') = U_i(s) - L_i(s)$).   However, if the packet is lost  due to channel erasure, the size of its Undelivered set  will remain  unchanged  (i.e., $U_i(s') = U_i(s)$).  Therefore, the expected cost of action $a$ on receiver $R_i$, targeted by its next needed packet, can be expressed as:
         \begin{align}\label{eqn:nextneeded}
         \bar{c}_i(s,a| R_i \in \mathcal{T}_{\rho}(a)) = (U_i (s) - L_i(s)) \times (1- \epsilon_i) + U_i(s) \times \epsilon_i = U_i(s) - L_i(s) \times (1-\epsilon_i).\nonumber
         \end{align}

   \item  Consider receiver $R_i$ either has been targeted by one of its needed packets or has not been targeted in this transmission, i.e., $R_i \in \mathcal{M}_w\setminus\mathcal{T}_{\rho}(a)$. Under both packet reception and loss scenarios,  the size of its Undelivered set will remain unchanged (i.e., $U_i(s') = U_i(s)$). Therefore, the expected cost of action $a $ on receiver $R_i$, either  targeted by one of its  needed packets or ignored in this transmission, can be expressed as: $\bar{c}_i(s,a| R_i \in \mathcal{M}_w \setminus \mathcal{T}_{\rho}(a)) =  U_i(s)$.

   \end{itemize}

 Having defined the expected  cost of  action $a = \kappa(s) \in \mathcal{A}(s)$ on each receiver $R_i \in \mathcal M_w(s)$, the total expected cost of   action $a$ over all receivers in $\mathcal M_w(s)$  can be expressed as:
  \begin{align}
  \bar{c}(s,a) &= \sum_{R_i \in \mathcal{T}_{\rho}(a)} U_i(s) - (L_i(s) \times (1-\epsilon_i)) + \sum_{R_i \in \mathcal{M}_w \setminus \mathcal{T}_{\rho}(a)} U_i(s). 
  \end{align}
\end{enumerate}

\vspace{-5mm}
\subsection{Policies of the Formulated SSP Problem} \label{SSPproperties}
An SSP policy $\pi=[\pi(s)]$ is a mapping from  $\mathcal{S}\rightarrow\mathcal{A}$ that associates an action to each of the states.  The algorithms solving SSP problems define a value function $V_{\pi}(s)$ as the expected cumulative cost until completion, when the system starts at state $s$ and follows policy $\pi$. It is recursively expressed $\forall s\in\mathcal{S}$ as \cite{puterman2009markov}:
\begin{align}
V_{\pi}(s) &=   \bar{c}(s,\pi(s)) + \sum_{s' \in \mathcal{S}(s,a)} \mathcal{P}_{\pi(s)}(s,s')V_{\pi}(s'),
\end{align}
where, $\mathcal{S}(s,a)$ is the set of successor states to state $s$  when action $a$ is taken following policy $\pi(s)$ (i.e., $ \mathcal{S}(s,a) = \{s'|\mathcal{P}_{a}(s,s') > 0 \} $). The optimal policy $\pi^*(s)$ at  state $s$ is the one that minimizes the number of  undelivered packets to all receivers over all transmissions until completion, and  can be expressed $\forall s \in \mathcal{S}$ as:
\begin{align} \label{eqn:optimal}
\pi^*(s) &=  \arg\min_{a \in \mathcal{A}(s)} \left\{ \bar{c}(s,a) + \sum_{s' \in \mathcal{S}(s,a)} \mathcal{P}_{a}(s,s')V_{\pi^*}(s')  \right\}.
\end{align}
According to \eqref{eqn:optimal}, the optimal action at state $s$  depends on the immediate cost as well as the expectation of the value functions of the successor states. Similarly, we  state that the policies that can  efficiently reduce the  number of  undelivered packets to all receivers over all transmissions  should focus, at any  state $s$, on both:
\begin{itemize}
\item  \emph{Immediate cost:} Bringing the  Undelivered vector $\mathbf{u}(s)$ close to the absorbing state vector $\mathbf{u}(s_a)$. In other words, targeting receivers with their next needed packets.
\item  \emph{Value functions of the successor states}: Increasing the sizes of the Potential sets  in the  successor states  of state $s$. In other words, increasing the number of  decoded packets at the receivers (since all these decoded packets will be delivered in  future transmissions upon receiving all their preceding missing packets).
\end{itemize}
\vspace{-5mm}
\subsection{SSP Solution Complexity} \label{complexity}
The optimal policy of the formulated SSP problem can be computed using the policy iteration algorithm with  complexity $O(|\mathcal{S}|^3 + |\mathcal{S}|^2|\mathcal{A}|)$  \cite{puterman2009markov}.  Based on the sizes of  $\mathcal{S}$  and  $\mathcal{A}(s)$  of the formulated  SSP problem,  we conclude that the policy iteration algorithm quickly leads to computational intractability even for systems with moderate numbers of  receivers  and packets.

\vspace{-5mm}
\section{Guidelines for  Efficient Packet Selection Policies}\label{guidelines}
In this section,  we will  explore  the  in-order packet delivery aspect of the formulated  SSP problem and  draw  guidelines for the packet selection policies that can efficiently reduce the number of undelivered packets to all receivers over all transmissions until completion.
\vspace{-5mm}
\subsection{Effect of  Orders of the Missing  Packets at their Respective Receivers on the Coding Decisions} \label{sec:packet}

The in-order packet delivery  constraint requires the sender to target the receivers with their next needed  packets. In SSP terms,  this can be translated  as selecting a policy at the sender  that quickly reduces    the number  of undelivered packets  at the receivers  and results in a low cumulative  cost. Therefore, an efficient coding decision needs to prioritize the missing packets according to  their orders at their respective receivers so that the received packets are immediately  delivered, if the receivers are  targeted by the next needed packets, or   quickly delivered in future transmissions, if the receivers are targeted by other needed packets.

To systematically  capture such  packet prioritization, given an SFM at time slot $t$, we first arrange  the missing packets  of each receiver in non-decreasing order of the  packet indices.  For instance, given the SFM in \eqref{largeSFM1}, missing packets are arranged as $\{P_1,P_3\}$ and $\{P_3,P_4,P_6\}$ for receivers $R_1$ and $R_2$, respectively. We  then classify all  missing packets into groups  such that    the first missing packets of all receivers (i.e., the next needed packets) belong to  Group 1, the second missing packets of all receivers belong to  Group 2 and so on. Therefore,  the   number of groups for a given SFM can be defined as, $D = \max_{i \in \mathcal{M}_w}\{W_i\}$. Now, we list all groups in non-decreasing order of the group numbers. This means   Group 1 containing the next needed packets is placed first in the list.  Having defined the  groups and their orders, we finally set the priority of a missing packet belonging to a group as $D - d_{ij} +1$, where $d_{ij}$ is the   $d-$th order group among all $D$ groups that contains missing packet $P_j$ of receiver $R_i$. 

\begin{examples} Let us consider the SFM in  \eqref{largeSFM1}, where  the size of the largest Wants set is 3  and thus, the number of groups is $D = 3$. Vertices $v_{1,1}, v_{2,3}$ (next needed packets)\footnotemark \footnotetext{Vertex $v_{2,3}$ represents missing packet  $P_3$ at receiver $R_2$ in IDNC graph $\mathcal{G}$ constructed from the SFM in \eqref{largeSFM1}.}  belong to the first group,  vertices $v_{1,3},$ $v_{2,4}$ belong to the second group  and  vertex $v_{2,6}$ belongs to the third group. The prioritization of each vertex belonging to the first, second and third  groups can be calculated as, $3$  $(3 - 1 +1 = 3), 2$ and $1$, respectively.
\end{examples}

In fact, the next needed packets of all receivers  have the same prioritization as they belong to the same group, and the next needed packet of any receiver has a higher prioritization than  other needed packets of all receivers since it belongs to the first group. These observations also hold for other needed packets.

\vspace{-5mm}
\subsection{Effect of  Previously Decoded but Undelivered Packets on the Coding Decisions}\label{sec:receiver}

Here, we explore the aspect of  delivering a burst of in-order decoded packets upon decoding a missing packet and thus,  quickly moving the Undelivered set to the completion state (i.e., $U_i = 0, \forall R_i\in \mathcal{M}$). Since the cost in the SSP formulation depends on  the size of the Undelivered sets, a quick reduction of such sets results in  a low cumulative cost. In fact, given an SFM at  time slot $t$, it is possible that there are previously decoded packets at a receiver and these decoded packets cannot be delivered  because of missing at least one
of their preceding packets. To make efficient coding decisions, the sender  needs to take into account the effect of decoding  a missing packet on delivering  a burst of previously decoded packets. 

\begin{definitions}
\emph{At any given time slot $t$, the packet delivery rate for  receiver $R_i$ is defined by, $\frac{U_i}{W_i}$, the average rate at which the packets are  delivered to the receiver upon decoding a missing packet.\footnotemark}
\footnotetext{This definition  represents  the average number of delivered packets to a receiver over decoding all of its missing packets. Therefore, after decoding a missing packet at a receiver in a transmission,  the number of delivered packets will  not necessarily be equal to its delivery rate.}
\end{definitions}


Given the SFM in \eqref{largeSFM1}, the packet delivery rate for receiver $R_1$ is $\frac{6}{2} = 3$. This means  on average three packets are delivered to receiver $R_1$ upon decoding a missing packet.
In fact, at any visited state $s$, the delivery rate exploits the  status of previously decoded but undelivered  packets at a receiver and  captures the rate at which the Undelivered set reaches  its completion state of the SSP formulation. Having discussed the  packet and receiver prioritization in Sections \ref{sec:packet} and \ref{sec:receiver} separately, we define the prioritization of packet $P_j$ for receiver $R_i$ as,  $\psi_{ij} = (\frac{U_i}{W_i})^{\alpha}(D - d_{ij} +1)$, where $\alpha \in \{0,1,2,3,...\}$ is a biasing factor that  allows to select different importance of the delivery rate in making coding decisions.

\vspace{-5mm}
\subsection{Effect of  Channel Erasures  on the Coding Decisions}
For erasure channels,  the impact of  erasures should be reflected on the coding decisions. Therefore,  consistent with a low cumulative cost in the SSP formulation, we  give  a high priority of service  to  a receiver having a high packet reception probability compared to other receivers having low packet reception probabilities. To implement such channel prioritization, we define channel-aware delivery rate for receiver $R_i$ as, $(1-\epsilon_i)\left(\frac{U_i}{W_i}\right)$. Indeed, a receiver having good channel condition has high probability of receiving  and delivering  of its undelivered packets. Finally, we redefine the  prioritization of packet $P_j$ for receiver $R_i$ as:
\begin{equation} \label{channelweighted1}
\tilde{\psi}_{ij} = (1-\epsilon_i)\left(\frac{U_i}{W_i}\right)^{\alpha}(D - d_{ij} +1).
\end{equation}

\vspace{-5mm}
\section{Heuristic Algorithm for Packet Selection} \label{heuristic}
In this section, we  design a simple  heuristic algorithm that reduces the  number of undelivered packets to all receivers over all transmissions until completion. At any visited state $s$,  the heuristic algorithm selects a maximal clique $\kappa^*$ based on a greedy  maximum weight vertex search over the IDNC graph $\mathcal{G}(s)$.  To define the vertices' weights, we first define $e_{ij,kl}$ as the adjacency indicator of vertices $v_{ij}$ and $v_{kl}$ in $\mathcal{G}(s)$ such that:  $e_{ij,kl} = 1,$ if $v_{ij}$ is connected to $v_{kl}$,  and  $e_{ij,kl} = 0,$ otherwise.
We then define the weighted degree $\Theta_{ij}(s)$ of  vertex $v_{ij}$  as: $\Theta_{ij}(s) = \sum_{v_{kl} \in \mathcal{G}(s)} e_{ij,kl} \tilde{\psi}_{kl}(s)$, where $\tilde{\psi}_{kl}(s)$ is the prioritization of packet $P_l$ for receiver $R_k$ as defined in \eqref{channelweighted1}.   We finally define  the weight  of vertex $v_{ij}$ as:
\begin{equation}\label{w:weight}
w_{ij}(s) = \tilde{\psi}_{ij}(s) \Theta_{ij}(s) = \left\{(1-\epsilon_i)\left(\frac{U_i}{W_i}\right)^{\alpha} (D - d_{ij} +1)\right\} \Theta_{ij}(s).
\end{equation}

Having defined the vertices' weights,  the heuristic algorithm evolves  as follows. At Step 0,  there are no  vertices in the selected maximal clique $\kappa^*$. At Step 1, the algorithm selects the vertex $v_{ij}^*$ that has the maximum weight $w_{ij}^{\mathcal{G}(s)}$ and adds it to $\kappa^*$ (i.e., $\kappa^* = \{ v_{ij}^*\}$). After Step 1, the algorithm extracts the subgraph $\mathcal{G}(\kappa^*)$ of vertices in $\mathcal{G}$ that are adjacent to all previously selected vertices in $\kappa^*$. It then recomputes the weights of the vertices in subgraph $\mathcal{G}(\kappa^*)$. At Step 2, the algorithm selects vertex $v_{kl}^*$ that has the maximum weight $w_{kl}^{\mathcal{G}(\kappa^*)}$ and adds it to $\kappa^*$ (i.e., $\kappa^* = \{\kappa^*, v_{kl}^*\}$). This process is repeated until no further vertices are adjacent to all the  vertices  in $\kappa^*$. Once the maximal clique is selected, the sender forms  a coded packet by XORing the source packets identified by the vertices in $\kappa^*$. We refer to this algorithm as \emph{maximum weight vertex search} (`MWVS') algorithm. The  complexity of  the  MWVS algorithm is $O(M^2N)$ since it requires weight computations for the $O(MN)$ vertices  in each step and  a maximal clique can have at most $M$ vertices. 

\vspace{-5mm}
\section{Simulation Results} \label{results}
In this section, we  present the simulation results comparing  the performance of the policy iteration (`PI') algorithm that solves the formulated SSP problem  and the proposed  MWVS algorithm to the following   algorithms. (\textbf{A1}): Interrelated priority encoding   (`IPE-Two') algorithm, proposed in  \cite{wanginstantly}, that adopts a two-phase transmission setting  and  reduces completion time  while respecting in-order packet delivery.  (\textbf{A2}): Modified interrelated priority encoding   (`IPE-Single') algorithm  that represents  a single-phase transmission version (as proposed this paper) of the  packet selection algorithm proposed in  \cite{wanginstantly}.  (\textbf{A3}): Completion time   (`CT') reduction algorithm \cite{sorour2012completion}  that ignores in-order packet delivery. (\textbf{A4}):  The  (`Mixed') algorithm \cite{aboutorabenabling}  that balances between reducing  completion time and servicing a large number of receivers with any new packet in each transmission. (\textbf{A5}): The  (`Max-Clique') algorithm \cite{le2013instantly}  that services a large number of receivers with any new packet in each transmission. The main characteristics of these algorithms are summarized in Table II.


%

\begin{figure}
\centering
\includegraphics[scale=0.8]{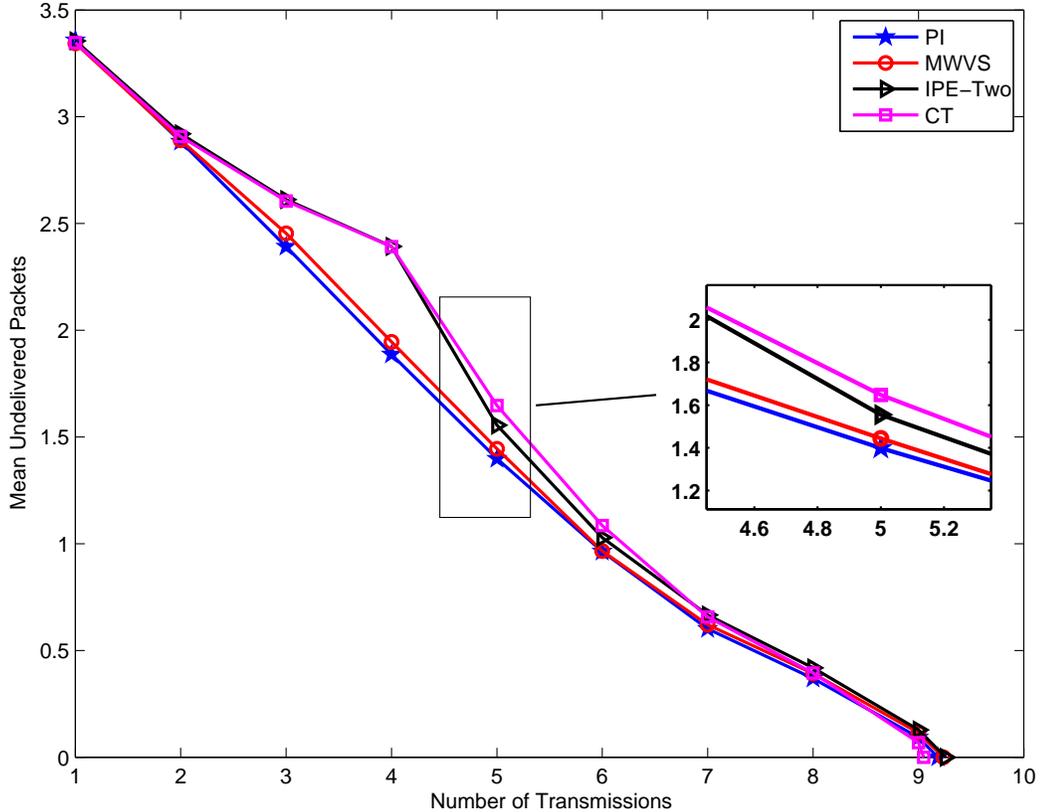}\\
\caption{Mean undelivered packets after different number of  transmissions} \label{fig:trans}
\end{figure}

For our proposed MWVS algorithm, we use biasing factor   $\alpha = 2$  in all scenarios. However, other biasing factors  are also possible. Fig. \ref{fig:trans} depicts the  mean undelivered packets after different number of transmissions achieved by different algorithms (for $M = N = 4$ and $\epsilon_1 = 0.2, \epsilon_2 = 0.3, \epsilon_3 = 0.4, \epsilon_4 = 0.5 $).\footnotemark \footnotetext{As discussed in Section \ref{complexity}, the complexity of the policy iteration (PI) algorithm scales with $|\mathcal{S}|$, which is  $2^{16} $  even for the considered system with $M = N = 4$. Note that the simulation results are the  average  based on over 2000 runs.}  The  mean undelivered packets after time slot $t$ is defined as the average  number of undelivered packets over all receivers. This can be expressed as: $\frac{\sum_{i \in \mathcal{M}} \hat{U}_{i,t}}{M}$, where  $\hat{U}_{i,t}$ is the number of undelivered packets to receiver $R_i$ after time slot $t$.   From this figure,  we can see that the performance  of the MWVS algorithm closely follows   the PI algorithm, the solution of the SSP formulation. Indeed, the MWVS algorithm is designed based on the guidelines derived from the in-order packet delivery aspect of the SSP formulation. This figure also shows that the performance of the  IPE-Two and CT algorithms substantially deviates from that of the PI algorithm, especially in the initial  four transmissions when these algorithms send four uncoded packets following the two-phase transmission setting, as discussed in Section \ref{sec:initial}.

\begin{figure}
\begin{center} \subfigure[]{\label{fig:rx}\includegraphics[width=11cm,height=7.2cm]{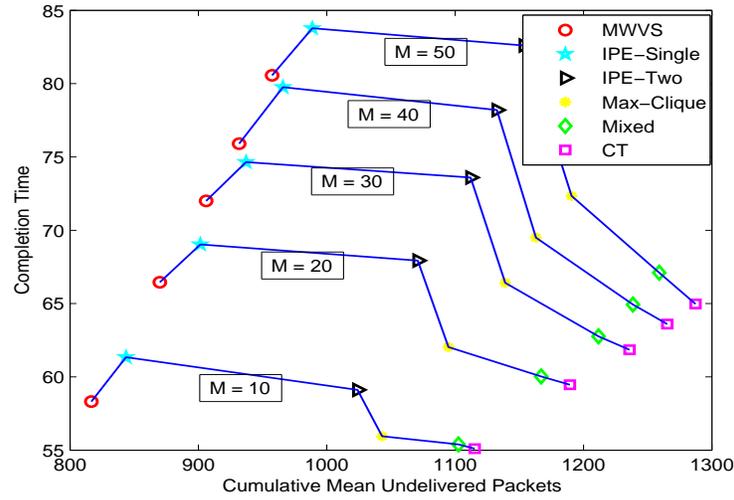}}
\subfigure[]{\label{fig:pk}\includegraphics[width=11cm,height=7.2cm]{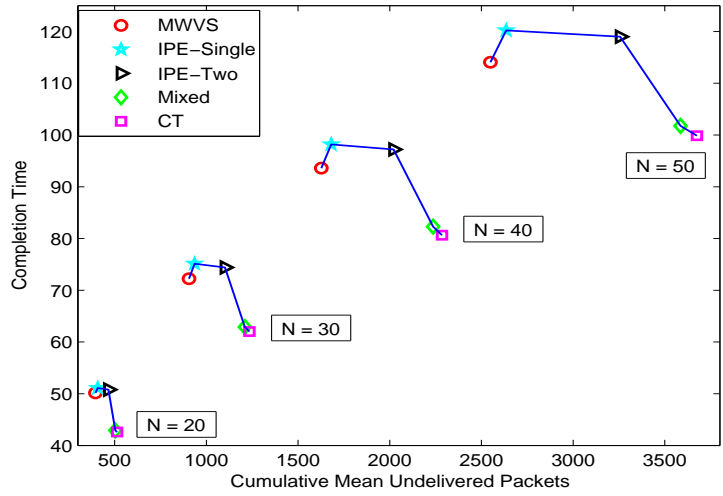}}
\subfigure[]{\label{fig:erasure}\includegraphics[width=11cm,height=7.2cm]{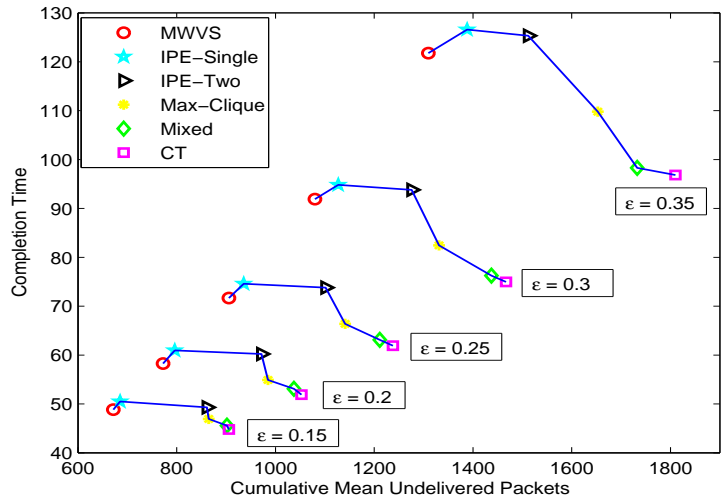}}
\end{center}
\caption{Completion time  versus   cumulative mean undelivered packets for   (a) different number of receivers $M$, (b)  different number of packets $N$, (c)  different average erasure probabilities $\epsilon$.}
\label{fig:second}
\end{figure}

Figs. \ref{fig:rx},  \ref{fig:pk} and  \ref{fig:erasure}  depict the completion time  and the  cumulative mean undelivered packets performances of different algorithms for different  number of receivers $M$ (for $N =30$ and $\epsilon = 0.25$),  different number of packets $N$ (for $M =30$ and $\epsilon = 0.25$) and different average erasure probabilities $\epsilon$ (for $M =30$ and $N = 30$),  respectively.\footnotemark \footnotetext{When average erasure probability $\epsilon = 0.25$, the  erasure
probabilities of different receivers are in the range $[0.05,0.45]$.}  The  cumulative mean undelivered packets
is calculated by summing  the mean undelivered packets over all transmissions until completion. This can be expressed as: $\sum_{t=1}^T \frac{\sum_{i \in \mathcal{M}} \hat{U}_{i,t}}{M}$, where $T$ is the completion time.   From all these figures, we can draw the following observations:
\begin{itemize}
\item Our proposed channel-aware MWVS algorithm outperforms the channel-unaware  IPE-Single and IPE-Two  algorithms in terms of the cumulative mean undelivered packets for all comparison parameters ($M, N$, $\epsilon$).
    In fact,  MWVS algorithm employs the IDNC graph  to exploit  all  feasible packet combinations  and prioritizes a packet  by capturing the effect of decoding this packet on  quickly delivering a burst of in-order decoded packets. Note that the significant performance degradation  of the IPE-Two algorithm is because of adopting the two-phase transmission setting with the aim  of reducing the completion time.
\item  The performance of the Max-Clique, Mixed and CT algorithms substantially deteriorates  compared to MWVS algorithm in terms of cumulative mean undelivered packets. Unlike the MWVS  algorithm, Max-Clique, Mixed and CT algorithms adopt the two-phase transmission setting and ignore the aspect of in-order packet delivery  in making coding decisions.
\item Our proposed MWVS algorithm outperforms the IPE-Single and IPE-Two  algorithms in terms of completion time for all comparison parameters ($M, N$,  $\epsilon$).  However,  as expected, CT algorithm achieves the best completion time performance because of  adopting the two-phase  transmission setting and making coding decisions with the specific and single aim of  reducing the completion time.
\end{itemize}

\vspace{-5mm}
\section{Conclusion} \label{conclusion}
In this paper, we studied  in-order packet delivery  in  IDNC systems for  wireless broadcast networks.  We formulated the problem of  minimizing the  number of undelivered packets to all receivers over all transmissions until completion as  an SSP problem, and  showed that finding the optimal packet selection policy using SSP is computationally complex.  However, exploiting the in-order packet delivery aspect of the SSP formulation, we  drew  guidelines for   efficient  packet selection policies and  designed  a  heuristic packet selection algorithm.   Simulation results showed  that  our proposed   algorithm  provides quicker packet delivery  to the receivers compared to the  existing algorithms.

\vspace{-5mm}
\bibliographystyle{IEEEtran}
\bibliography{ref}

\newpage

\begin{table}
 \caption{Main notations and their descriptions}
 \centering
    \begin{tabular}{|c| p{15.5cm}|}
    \hline
    \textit{Notation} &  \textit{Description} \\ \hline
    $\mathcal{N}$ & The set of $N$ packets \\ \hline
    $P_j$ &  The j-th packet in $\mathcal{N}$\\ \hline
    $\mathcal{M}$ & The set of $M$ receivers \\ \hline
    $R_i$ & The i-th receiver in $\mathcal{M}$\\ \hline
    $\mathcal{M}_w$ & The set of receivers with non-empty Wants sets \\ \hline
    $\mathbf{F}$ & $M \times N$ state feedback matrix (SFM)\\ \hline
    $\epsilon_i$ & Channel erasure probability experienced by receiver $R_i$ \\ \hline
    $\mathcal{H}_i$ & (Has set) The set of packets successfully decoded by  receiver $R_i$  \\ \hline
    $\mathcal{W}_i$ & (Wants set) The set of missing packets at receiver $R_i$ \\ \hline
    $\mathcal{U}_i$ & (Undelivered set) The set of undelivered packets to receiver $R_i$ \\ \hline
    $\mathcal{L}_i$ & (Potential set) The set of  packets that can be delivered to receiver $R_i$ upon decoding the next needed packet \\ \hline
    $\mathcal{G}$ & An IDNC graph constructed from an SFM \\ \hline
    $v_{ij}$ & A  vertex in an IDNC graph induced by  missing packet $P_j$ at receiver $R_i$\\ \hline
    $\kappa$ & A maximal clique in an IDNC graph $\mathcal{G}$\\ \hline
    $\mathcal{T}_{\rho}(\kappa)$ & The set of  receivers which are  targeted by their next needed packets in  maximal clique $\kappa$\\ \hline
    $\mathcal{T}_{\sigma}(\kappa)$ & The set of  receivers which are  targeted by their other needed packets in  maximal clique $\kappa$\\ \hline
    $s$ & A state  in our SSP formulation $(s \in \mathcal{S})$ \\ \hline
     $s'$ & The successor state of state $s$ \\ \hline
    $a$ & An action is a maximal clique $\kappa$ in an IDNC graph $\mathcal{G}$ \\ \hline
    $D$ & Number of groups required to classify all missing packets  of all receivers \\ \hline
    $d_{ij}$ & The d-th order  group among all $D$ groups that contains packet $P_j$ of receiver $R_i$   \\ \hline
    $\tilde{\psi}_{ij}$ & The prioritization of packet $P_j$ for receiver $R_i$ (vertex $v_{ij}$)\\ \hline
    $\hat{U}_{i,t} $ &  Number of undelivered  packets to receiver $R_i$ after time slot $t$ \\ \hline
    \end{tabular}
\end{table}

\begin{table}
 \caption{Algorithms and their main characteristics}
 \centering
    \begin{tabular}{|l|p{6cm} |p{2.4cm}  |p{4cm} |}
    \hline
    \textit{Algorithm} &  \textit{Main objective} &  \textit{Transmission setting} &  \textit{Coding decisions based on packet delivery constraint} \\ \hline
    Policy Iteration & Quick packet delivery  &  Single-phase  & In-order \\ \hline
    MWVS & Quick packet delivery &  Single-phase  & In-order \\ \hline
    IPE-Two \cite{wanginstantly} & Completion time reduction  and respecting quick packet delivery &  Two-phase  & In-order \\ \hline
    IPE-Single & Completion time reduction and respecting quick packet delivery &  Single-phase  & In-order \\ \hline
    CT \cite{sorour2012completion}& Completion time reduction &  Two-phase  &Any-order \\ \hline
    Mixed \cite{aboutorabenabling}& Balancing between  completion time reduction and servicing a large number of receivers with any new packet in each transmission&  Two-phase  &Any-order \\ \hline
    Max-Clique \cite{le2013instantly}& Servicing a large  number of receivers with any new packet in each transmission&  Two-phase  & Any-order \\ \hline

    \end{tabular}
\end{table}

%

\end{document}